\documentstyle[12pt,epsfig]{article}
\def\slash#1{\setbox0=\hbox{$#1$}#1\hskip-\wd0\hbox to\wd0{\hss\sl/\/\hss}}
\begin{document}
\baselineskip=20 pt
\def\l{\lambda}
\def\m{\mu}
\def\L{\Lambda}
\def\bt{\beta}
\def\mphi{m_{\phi}}
\def\hphi{\hat{\phi}}
\def\vphi{\langle \phi \rangle}
\def\etamunu{\eta^{\mu\nu}}
\def\dmul{\partial_{\mu}}
\def\dnul{\partial_{\nu}}
\def\bea{\begin{eqnarray}}
\def\eea{\end{eqnarray}}
\def\bfl{\begin{flushleft}}
\def\efl{\end{flushleft}}
\def\bgt{\beta (g_t)}
\def\bgtmz{\beta (g_t(m_z))}
\def\blam{\beta (\lambda )}
\def\vphic{\vphi_{crit}}
\def\mphi{m_{\phi}}
\begin{center}

{\large \bf
  The effect of a light radion on the triviality \\

\vskip 0.04in

 bound on higgs mass. }

\end{center}

\vskip 10pT

\begin{center}
{\large\sl \bf{Uma Mahanta}~\footnote{E-mail: mahanta@mri.ernet.in}}
\vskip  5pT
{\it
Mehta Research Institute, \\
Chhatnag Road, Jhusi
Allahabad-211019, India .}\\
\end{center}


\begin{center}
{\large\sl  \bf{Prasanta Das}~\footnote{E-mail:pdas@mri.ernet.in}}
\vskip 5pT
{\it
Mehta Research Institute, \\
Chhatnag Road, Jhusi
Allahabad-211019, India .}\\
\end{center}

\centerline{\bf Abstarct}

\noindent In this paper we study how the triviality bound on higgs mass 
in the context of the SM is modified by a light stabilized radion of 
the Goldberger-Wise variety. Our approach is
inherently perturbative. Including the radion contribution to $\bt(\l)$ and
$\bt(g_t)$ to one loop we evolve the higgs self coupling $\l$ from the
cut off $\L(=\vphi)$ down to the EW scale $\mu_0 = v$. The triviality bound
is obtained by requiring that $\l(\L) = \sqrt{4 \pi}$ which is the
perturbative limit. We also study the effect of small changes in the UVBC
on the triviality bound both in the presence and absence of a light radion.

\newpage

\bfl
{\large {\bf {1.  Introduction}}}
\efl
\noindent In spite of a lot of searches the higgs particle of Standard 
Model (SM) has not been discovered so far. LEP2 searches based on the process
$e^+ e^- \rightarrow Z h$ have produced the lower bound $m_h > 113.2$
GeV for the SM higgs boson \cite{Hg1}.  The higgs particle also contributes to
EW radiative
corrections and this has been used to derive an upper bound
$m_h < 212$
GeV at 95\% C.L. \cite{Hg2}. Both these limits are of course
valid only in the SM.
There also exist several theoretical constraints on $m_h$ like the
triviality bound and vacuum stability bound. It is well known that the
renormalized $\phi^4$ theory in $3+1$ dimension cannot contain an 
interaction
term $\l \phi^4$. In other words the $\phi^4$ theory in $3+1$ dimension
must be trivial
(non interacting) if it is to be valid at all scales \cite{KW}.
This means that
in perturbation theory the running coupling $\l(\mu)$ diverges at some
finite value of $\mu$.  A similar effect occurs in the SM, modified to some
extent by gauge and Yukawa coupling of top quark \cite{MPP}.
Therefore if the scalar sector of the
SM is to be non trivial then the SM must be valid only upto some finite
scale $\L$.

Assuming  that the SM is an effective theory valid below some cut off
 scale $\L$, then
the running coupling $\l(\mu)$ must diverge at $\L$ or above it. This
condition gives us a $\L$ dependent upper bound on $m_h$. The triviality
bound on $m_h$ is usually derived assuming pure SM interactions. However, in
some new physics models one or more light dynamical fields could be present
besides the SM particles. If these new light fields couple to the higgs boson
with appreciable strength
then they could contribute to $\bt(\l)$ and thereby generate important
modifications to the triviality bound on $m_h$. In this paper we shall
consider the modifications to the triviality bound on $m_h$ that arises
from a light stabilized radion in the Randall-Sundrum(RS1) model \cite{RS}.
The radion will be assumed to be stabilized by the Goldberger-Wise
mechanism \cite{GW1} which is
known to produce a light stabilized radion. We shall compare the SM
predictions on triviality bound with those of the radion+SM system
for cut off scales between $500$ GeV and $10$ TeV. In our calculations we shall
set $\L = \vphi$ and assume that the KK modes of the graviton are heavier
than $\L$. We shall further assume that the heavy KK modes of the graviton
decouple from the light physics at the Electro-Weak(EW) scale.
Under this condition the
effect of the KK modes on light physics could be represented by a series of
higher dimensional gauge invariant operators made out of SM fields. The
operators will be suppressed by suitable inverse powers of $\L$. However, it
has been shown that the effect of such higher dimensional operators does not
change the triviality bound on $m_h$ significantly from the SM prediction
\cite{GW2}.
On the other hand we shall find that a light radion drastically changes the
triviality bound from its SM prediction for $\vphi$ less than about
1 TeV. In this paper we shall also study the effect of small changes in the
UVBC on the triviality bound both in the SM and radion+SM system.

\bfl
{\large {\bf {2. Determining the triviality bound for radion + SM system}}}
\efl
In order to determine the triviality bound for the radion+SM system
corresponding to some physical cut-off $\L$ by perturbative calculations
we need the beta function for the higgs self coupling $\l$ for the same.
The SM
contribution to $\bt(\l)$ in one loop is well known and is given by
\cite{BHL}.
\bea
{\beta_{SM} (\lambda )} ={1\over8\pi^2}\left[9\lambda^2 +
+\lambda (6 g_t^2-{9\over 2} g_2^2-{3\over 2}g_1^ 2)
-6 g_t^4+{3\over 16} (g_2^4+{1\over 2}(g_2^2+g_1^
2)^2)\right]
\eea
where $g_t$ stands for the Yukawa coupling of the top quark.

\noindent The radion contribution to $\bt(\l)$ to one loop can be shown 
to be given by \cite{DM1}
\bea
{\beta_{rad} (\lambda )} ={1\over8\pi^2}\left[{402 \lambda^2 v^2\over \vphi^2}
+ {144\lambda^2 v^4\over \vphi^4}
+{5\lambda \mphi^2 \over \vphi^2}\right]
\eea

\noindent The beta function of $\l$ for the radion + SM is therefore given by,
\bea
{\beta (\lambda )}  = \mu {d\lambda \over d\mu }={1\over
8\pi^2}\left[9\lambda^2 +
{402 \lambda^2 v^2\over \vphi^2}+ {144\lambda^2 v^4\over \vphi^4} +
{5\lambda \mphi^2 \over \vphi^2}
+\lambda (6 g_y^2-{9\over 2} g_2^2-{3\over 2}g_1^ 2)\right]
\nonumber \\
+\frac{1}{8 \pi^2} \left[-6 g_y^4+{3\over 16} (g_2^4+{1\over 2}(g_2^2+g_1^
2)^2)\right]
\eea

\noindent The expression for $\bt(\l)$ besides depending on radion 
mass $m_\phi$ and
radion vev $\vphi$ also depends on the top Yukawa coupling $g_t$ and the
EW gauge couplings $g_1$ and $g_2$. We need the beta functions of the later
couplings also for the radion+SM system in order to evolve $\l$ from the
cut off scale $\L$ down to the EW scale. Since the radion is an EW singlet,
the beta functions for $g_1$ and $g_2$ in the radion+SM system are given by
the same expressions as in the SM. The radion however does couple to the
top quark quite strongly
 and contributes to $\bt(g_t)$. It can be shown that the beta
function of $g_t$ for the radion+SM system to one loop order is given by
\cite{DM2}
\bea
\bt(g_t) = \bt_{SM}(g_t) + \bt_{rad}(g_t)
=\frac{g_t}{16 \pi^2}
\left[\frac{9}{2} g_t^2 - 8 g_3^2 - \frac{9}{4} g_2^2 -
\frac{17}{12} g_1^2\right]
\nonumber \\
+ \frac{g_t}{16 \pi^2 \vphi^2}\left[ 4 m_\phi^2
+ \frac{31}{2} g_t^2 v^2 + 9\l v^2 \right]
\eea

\noindent The radion contribution is given by those terms that vanish 
in the limit
$\vphi \rightarrow \infty$. With the above expressions for $\bt(\l)$ and
$\bt(g_t)$, we are now in a position to evolve $\l$ from the cut-off
scale $\L$ down to the EW scale. Although the choice of the cut off
$\L$ for the SM is somewhat arbitrary,
 for the radion+SM system in RS1
model a  natural cut off is provided by $\vphi$. For a viable solution to the
hierarchy problem the radion must be so stabilized that $\vphi$ lies between
few hundreds of GeV to a few TeV.

\noindent In this paper we shall determine the triviality constraint on 
$m_h$ for the
radion+SM system for cut-offs $\vphi$ lying between $500$ GeV and
$10$ TeV and compare
them with the corresponding SM predictions. In perturbation theory the
triviality bound can be determined from the assumption that the running
coupling $\l(\mu)$ blows up at the cut off. So to determine the triviality
bound perturbatively we shall work with the UVBC $\l (\vphi )=\sqrt {4\pi}$
which is the perturbative limit.
Further to
study the variation of the triviality bound with changes in the UVBC we
shall also determine the triviality bound for a different UVBC namely
$\l(\vphi) = 2.0$.
In Figure 1 we have plotted the triviality bound on
$m_h$ for cut offs $\vphi$ lying between $500$ GeV and $10$ TeV for the
 UVBC $\l(\vphi) = \sqrt{4 \pi}$.
Figure 2 is a repeat of Figure 1 but for the UVBC $\l(\vphi) = 2.0$.

\noindent We would like to note  the {\it following points}:
\begin{itemize}
\item{} The triviality bound on $m_h$ changes quite significantly under
the change in UVBC that we have made both in the SM and radion+SM system.
But the qualitative nature of the variation of triviality bound with
$\vphi$ remanis the same for both the SM and radion+SM system.

\vspace*{0.1in}
\item{} The triviality constraint increases monotonically with decreasing
cut off in the SM. This fact can be easily understood because with
decreasing cut off the running coupling $\l(\mu)$ gets smaller and smaller
momentum interval to run down from its value at the cut off.
On the other hand for
the radion+SM, for large enough cut offss the triviality bound increases with
decreasing cut off which is expected since the radion decouples
from the SM for large enough $\vphi$.
 However for cut offs smaller than 3 TeV the triviality
bound decreases with decreasing cut off. This behaviour is due to the fact
that for the radion+SM as the cut off $\vphi$ decreases the radion
contribution to $\bt(\l)$ increases. As a result for small enough $\vphi $
 although the running interval
shrinks with decreasing cut off, the beta function becomes much stronger
and causes $\l(\mu)$ to fall off very fast. The net result therefore
is a
decreasing triviality bound with decreasing cut off instead of an
increasing one.
\end{itemize}

\noindent In this paper we have assumed that the KK modes of the graviton are
several times heavier than the compactification scale $\vphi$ and that
they decouple from the light SM physics. In the RS1 model even the lightest
KK mode of the graviton turns out to be several times heavier than the
compactification scale $\vphi$. It is therefore reasonable to assume
that the KK graviton modes decouple from the SM.

\bfl
{\bf Conclusion}
\efl
In conclusion in this paper we have studied how a light
stabilized radion in the RS1 model affects the triviality bound on higgs
mass for varying cut offs between $500$ GeV and $10$ TeV. We have also studied
the effect of changing the UVBC i.e. the value of $\l(\vphi)$ on the
triviality bound. We find that for $\vphi$ greater than 3 TeV, the triviality
bound increases with decreasing cut off for the radion+SM system like in SM.
However for $\vphi$ less than 3 TeV the triviality bound decreases with
decreasing cut offs in sharp contrast to its SM behaviour where the triviality
bound increases monotonically with decreasing cut off.

\bfl
{\bf Acknowledgement}
\efl
We would like to thank Dr. B. Ananthanarayan for hepling us with his
code for solving the RG eqns for  $\l$ and $g_t$.

\newpage
\vspace*{-0.3in}
\begin{figure}[htb]
\begin{center}
\vspace*{0.5in}
\hspace*{-0.9in}
\epsfig{file=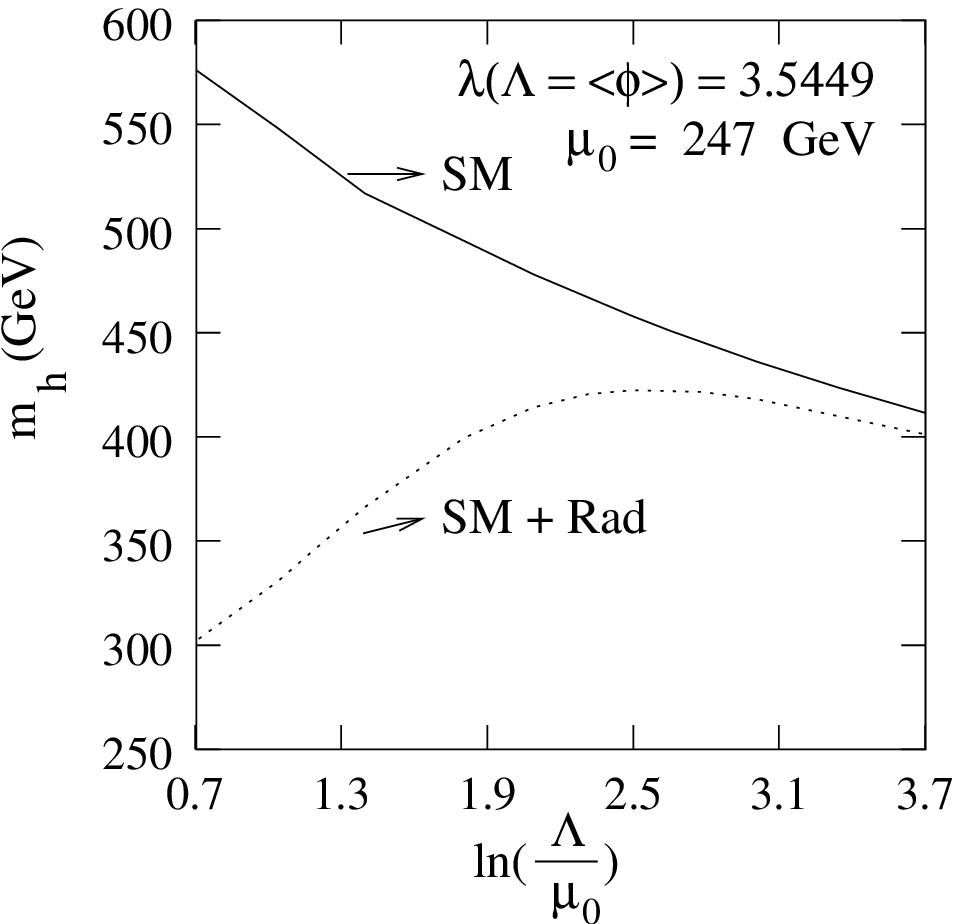}
\end{center}
\vspace*{-0.2in}
{\caption{\it { Triviality bound on $m_h$ in the radion+SM system correspondig
to $\l(\vphi) = \sqrt{4 \pi}$.}}}
\end{figure}
\newpage
\vspace*{-0.3in}
\begin{figure}[htb]
\begin{center}
\vspace*{0.5in}
\hspace*{-0.9in}
\epsfig{file=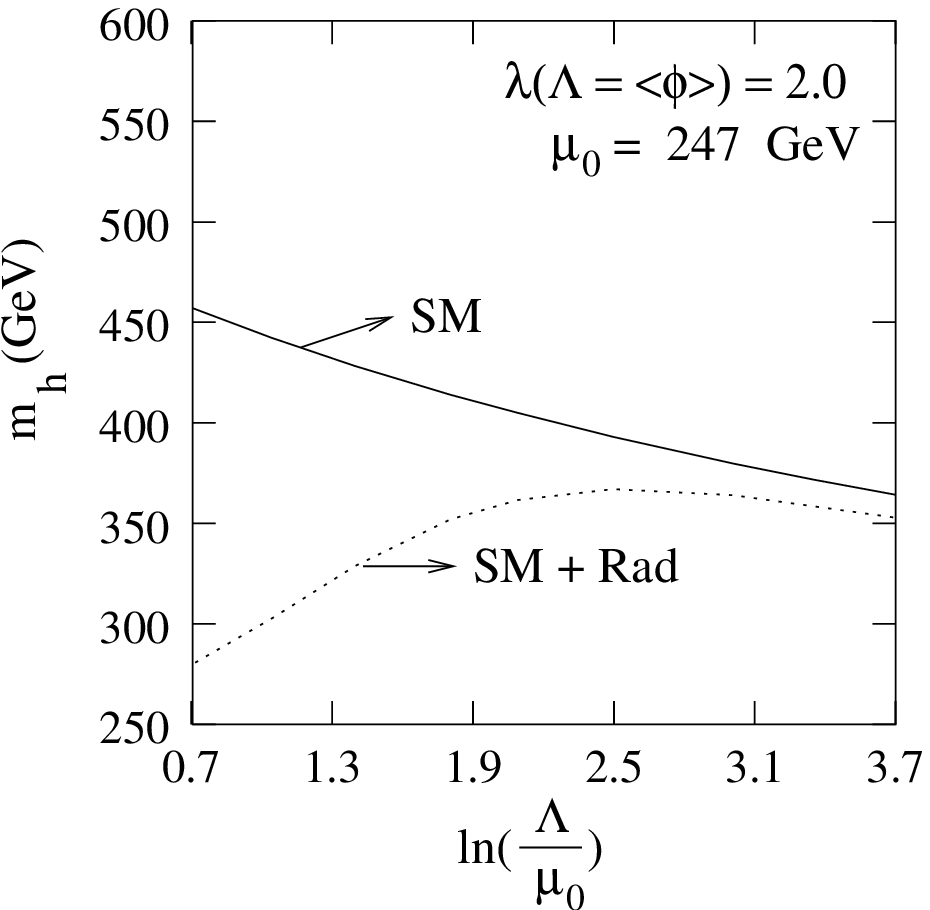}
\end{center}
\vspace*{-0.2in}
{\caption{\it { Triviality bound on $m_h$ in the radion+SM system correspondig
to $\l(\vphi) = 2.0$.}}}
\end{figure}

\end{document}